_______________________________________________________________________________

\documentstyle[preprint,aps]{revtex}
\begin{document}
\draft
\title{
EVIDENCE FOR  COLOR FLUCTUATIONS\\
IN HADRONS FROM COHERENT NUCLEAR DIFFRACTION}
\author{L. Frankfurt}
\address{
Tel Aviv University, Ramat Aviv, Israel\\
also at Institute of Nuclear Physics, St. Petersburg, Russia}
\author{G.A. Miller}
\address{Department of Physics, FM-15\\
University of Washington, Seattle, Washington 98195, USA}
\author{
M. Strikman}
\address{
Department of Physics\\
Pennsylvania State University, University Park, PA 16801, USA\\
also at Institute of Nuclear Physics, St. Petersburg, Russia}
\date{today}
\maketitle
\begin{abstract}
A QCD-based treatment of projectile size fluctuations is used to compute
inelastic diffractive cross sections $\sigma_{diff}$
for   coherent hadron-nuclear processes. We find that
fluctuations near the average size give the major contribution to the
cross section
with $ \le few \%$ contribution from small size configurations.
 The computed values of $\sigma_{diff}$ are consistent with
the limited available data.
The importance of
coherent diffraction studies for a wide range of projectiles
for  high  energy  Fermilab fixed target experiments is emphasized. The
implications of these significant color fluctuations
for relativistic heavy ion collisions
are discussed.

\end{abstract}
\newpage

We demonstrate that  color
fluctuations in the projectile
 wave function play an important role
 in high-energy ($E_h$) hadron (h)-nuclear reactions.
 This is done by studying the cross section $\sigma_{diff}$
for inelastic
 coherent nuclear diffraction which
 would vanish
 without such fluctuations.

Not much data for inelastic
 coherent nuclear diffraction exists. But
making the necessary measurements   seems
 much more feasible now than 10-15 years ago, due
to the development of micro-vertex detectors \cite{potter} in which
the target is the detector itself. Using these detectors   can insure
that the nucleus is left in its ground state in the diffractive process.
Furthermore, measuring  the coherent
nucleon-nuclear diffractive cross section
will ultimately
allow the computation
of fluctuations of various observables available in heavy
ion collisions \cite{HBB91,BF92}. This means that new experiments, which
can be done during the 1995 Fermilab fixed target run, have
gained a new urgency.

We begin our analysis with a general discussion.
 It has long been
known that the
 time scale, given by the uncertainty principle,
  for quantum fluctuations from a hadronic state h into a state X is
${2E_h\over{m^2_X-m^2_h}}$. Such
fluctuations are inhibited at
 large enough energies, so that
one may treat the hadron
 as frozen in its initial configuration \cite{Mandelstam}.
 The natural approach to describe such collisions is
the scattering-eigenstates formalism,
introduced by  Feinberg and Pomeranchuk \cite{FP56} and Good
and Walker \cite {GW60}. This accounts for the high energy coherence
effects:
 the projectile can be treated as a coherent superposition of
scattering eigenstates, each with an eigenvalue $\sigma$. The probability that
a given configuration interacts with a nucleon with a total cross section
$\sigma$ is $P(\sigma)$.  It is possible to reinterpret
$P(\sigma)$
by relating the size of a given configuration with its cross section (forward
scattering amplitude) in a monotonic fashion; see e.g.
\cite{Pum82,FS88,FS91b,fms92} and the references therein.

The scattering-eigenstates  formalism accounts for the effect that
different components of a hadron have different interactions with the
target.  For example:
(1) In color transparency physics a
 configuration with a small size has  small interactions.
(2) Nucleonic
configurations with and without a pionic cloud
 have different interactions with a target.
(3) In string
models of hadrons  the transverse size is approximately
proportional to the sine of the angle between the momentum and string
directions.

In quantum mechanics a system fluctuates amongst its components.
Such fluctuations lead to
 different kinds of interactions.
Since color dynamics determines all the interactions,
 we use the term color fluctuations.

What is known about the $P(\sigma)$?
 The convenient procedure is to consider moments of
this quantity:$\langle \sigma^n\rangle =\int d\sigma\sigma^n P(\sigma)$.
The zero'th moment is unity, by
 conservation of probability, and the first is the total cross section.
The  analysis  of
 diffractive dissociation data
from the  nucleon
\cite{MP78} as well as  inelastic corrections to the total hadron-deuteron
cross section \cite{KL78} determined  the second   moment of
$P(\sigma)$ (see the summary in \cite{BBF92}) while
 diffractive dissociation data from the
deuteron targets determined  the third moment
 of $P(\sigma)$ for protons\cite{BBF92}.
 The functional form of $P(\sigma)$ was then
determined by
taking the
  behavior for small values of $\sigma$
 from QCD
\cite {FS91a,pion92} and also including the rapid decrease of
$P(\sigma)$ for large values of $\sigma$.  Thus it is
possible to  obtain
a realistic form of $P(\sigma)$ for a wide range of
$\sigma$ \cite{BBF92,pion92}.

 The ideas behind the formulae for
$\sigma_{diff}(A)$ were suggested   a long time
ago, see e.g. Refs. \cite{MP78,Gri69,Gri70,I} and Ref.\cite{BS83}
and references therein.   Ref.~\cite {Alberi} reviews
the attempts to describe data in terms of a few scattering eigenstates.
However,
a realistic  model, based on QCD,  for the cross section fluctuations
was missing.
For example,
preQCD models contained terms corresponding to a $\delta (\sigma)$ piece of
 $P(\sigma)$ \cite{MP78,BS83} which  QCD does not allow.
Here we provide an expression for
$\sigma_{diff}(A)$ in terms of the
independent information given by
$P(\sigma)$.
We start with the standard formula for  the
diffractive cross section in terms of the transition matrix
$\hat T$:
\begin{equation}
d\sigma_{diff}(A)\propto \sum_{\alpha,M_X^2} \delta^4(P_f-P_i)\cdots
\mid\langle A; \alpha,M_X^2 \mid {\hat T }\mid A,h\rangle\mid^2,
\end{equation}
where the ellipsis represents the standard phase space and flux factors.
The frozen approximation allows us to use completeness to sum over
 the diffractive excitations $\alpha,M_X^2$. Then the only hadronic
 information resides in the square of the hadronic wave function.
The key step, introduced  by Miettinen and Pumplin \cite{MP78}
 and revived in Refs.\cite{HBB91,BBF92,pion92} is to re-express the
integral over that squared wave function  in terms of an integral over
$\sigma$
that involves the independently determined probability
$P(\sigma)$.
 For coherent nuclear processes the scattering wave function
can be obtained using the optical potential, now also a function
 of the integration variable $\sigma$.
We
 consider values of
 A greater than about 10,
 so that  the t-dependence of
 the nuclear form factor is much more important than that
  of the hN diffractive  amplitude. Then the
coherent nuclear
diffractive cross section $\sigma_{diff}(A)$ can be expressed as:
\begin{equation}
\sigma_{diff}(A)=\int d^2B \left[ \int d\sigma
P(\sigma) \sum_n  \left[<h\mid F(\sigma,B) \mid n>^2 \right]-\left[\int d\sigma
P(\sigma) <h \mid F(\sigma,B) \mid h>
\right]^2\right],\qquad
\end{equation}
 where
$ F(\sigma,B) = 1 - e^{-{1\over 2}\sigma T(B)}$
and $T(B)=\int_{-\infty}^\infty \rho_A(B,Z)dz$.
Here the direction of the beam is $\hat Z$ and the distance between the
projectile and the nuclear center  is $\vec R=\vec B + Z\hat Z$.
Eq.(2) is similar to the one used in \cite{BS83} which did not
introduce  the notion of
 $P(\sigma)$.
The advantage of eq. (2) as compared to  the related equation of
Ref.~\cite{ZKL81} (for a  review see
\cite{K90})    within the two gluon exchange model
is that  we do not need to assume the validity
 of pQCD at average interquark distances in hadrons where $\alpha_S$ is large
(with an extra prescription for
dealing with  gauge noninvariant effects due to introduction
 of nonzero gluon mass), nor
 use  constituent quark model wave
functions instead of parton wave functions.

It is instructive to consider the extreme
 black disk  (bd) limit of eq. (2).  Then
the function $F(\sigma,B)$ is unity
  for positions inside the nucleus and  zero otherwise,
so that $\sigma_{diff}(A)$ vanishes!  In particular,
the black disk model gives
$\sigma^{bd}_{tot}=2 \pi R_A^2$, $\sigma^{bd}_{el}=\pi R_A^2$
and $\sigma^{bd}_{diff}(A)=0$.
But we shall see that  including the effects of color fluctuations  leads to
observable diffractive cross sections rapidly increasing with A, which are
consistent with
 existing measurements of semi-inclusive diffraction \cite{Ferbel1,Ferbel2}.
Another way to show that color fluctuations cause
$\sigma_{diff} (A)$
is to observe that taking
$P(\sigma)$ to be  a delta function, e.g. $P(\sigma)=
\delta (\sigma-\bar\sigma)$ gives
$\sigma_{diff}(A)=0$.

We first display results for the pion projectile  and
use three parametrizations of
$P(\sigma)$ of Blattel et al \cite{pion92}
of the form
$P(\sigma) =
 N(a,n) e^{{-(\sigma-\sigma_0)^n\over (\Omega\sigma_0)^n}}$.
 All of these distributions have approximately the
same value of
\begin{equation}
\omega_\sigma\equiv
\frac{\langle \sigma^2\rangle -
\langle \sigma\rangle^2}{\langle \sigma\rangle^2},
\end{equation}
 with $\omega_{\sigma}=0.4 ~-0.5$.
The resulting $\sigma_{diff}(A)$ are shown
in Fig. 1a.
Note that for each $P(\sigma)$, $\sigma_{diff}(A)$ varies as
$A^{1.05}$ for  $A \approx 16$ and as
 $A^{0.65}$ for large $A \approx 200$.

Next we examine the $\sigma_{diff}(A)$ that can be
observed in proton
scattering. The
current data indicate
\cite{BBF92}
that $\langle \sigma^2\rangle\approx 1.25
 \langle \sigma\rangle^2$
(for proton energies of about 400 GeV)
so we may expect interesting effects.
The results are shown in Fig.~1b for three versions of $P(\sigma)$ of
ref.~\cite{BBF92}. The parameterization for the proton
is $P(\sigma) =
 N(a,n) {\sigma/\sigma_0\over \sigma/\sigma_0+a}e^{-{(\sigma-\sigma_0)^n
\over (\Omega\sigma_0)^n}}$.
The  curves are obtained with n=2, n=6 and n=10.
We see that the shape of $P(\sigma)$ plays an important role in obtaining
the magnitude of $\sigma_{diff}(A)$. The A-dependence is of the approximate
form
A$^{0.80}$ for $A \sim 16$ and
A$^{0.4}$ for $A \sim 200$,
 which is smaller than
for the pion case because here $P(\sigma=0)=0$
 and the average value of $\sigma$ is larger.

At present there are no data available on the A-dependence of
 $\sigma_{diff}(A)$.
However the A-dependence of the
reaction $\pi^+  + A \rightarrow \pi^+  +\pi^+  +\pi^-  +A$
 was studied in \cite{Ferbel1} for $p_{\pi^+}$ = 200 GeV.
 This data integrated over the mass
 interval $0.8 \le M_{3\pi} \le 1.5 GeV$  (and
 corrected for the small
 contribution of Coulomb excitations) are shown
 in Fig.1a. In Fig.1b we present
 the data of \cite {Ferbel2}on $n+ A \rightarrow p\pi^- + A$
 for the the mass interval $ 1.35 \le M \le 1.45 GeV$  with the
Coulomb contribution subtracted using
 the analysis of \cite {Ferbel2}.
The A-dependence of both pion and
 nucleon semi-inclusive diffraction
 is reproduced well by our calculation.
A priori,  the A-dependence for a given semi-exclusive
channel could be different
 from that of $\sigma_{diff}$. However, if fluctuations near the
average value
 of $\sigma$ dominate then,
 as we discuss below,
the A-dependence given by  eq.2 is sensitive
 mainly to the value of $\sigma_{tot}$, see eq.4.
Our  numerical results (using $P(\sigma)$) for $\sigma_{diff}(A)$
are reasonably close to
the calculation of the preQCD model of \cite {BS83} because
similar
 values of $\omega_{\sigma}$ are used. But  Ref. \cite {BS83}
 showed that their calculation
 agrees with the only experimental data available for  the
inelastic diffractive  total cross section,  obtained using
emulsion targets.  Hence we also agree with these data.

The similarity of the  results for
 different models of $P(\sigma)$  suggests that fluctuations of $\sigma$  near
the average give the major
contributions. This similarity is even more apparent in Fig.2, where
we plot $\sigma_{diff}(A)/\omega_\sigma$. To test this idea we compute
 an approximate
diffractive
cross section $\sigma_{diff}^{appr}$ by  using a Taylor series about
$\sigma=\langle \sigma\rangle $ in the integrals
$\int d\sigma P(\sigma) f(\sigma)$.
The result for
$\sigma_{diff}^{appr}$ is
\begin{equation}
\sigma_{diff}^{appr}={\omega_\sigma\langle \sigma\rangle^2\over 4}\int d^2B
 T^2(B)
e^{-\langle \sigma\rangle T(B)},
\end{equation}
which is shown as the dashed line in Fig. 2.
We see that the approximation is         quantitatively accurate  for
$A < 50 $
 and qualitatively good for all values of A.
This is because for realistic $P(\sigma)$  the dominant
 contribution to the inelastic diffraction cross section
arises from impact parameters B near the nuclear
 surface where
$\langle \sigma\rangle T(B)$ is small.
As a result, the second cumulant  (dispersion
 of the cross section) dominates the  diffractive cross section.
This shows that
the A dependence is mainly determined by the value of $\langle \sigma\rangle$.
The deviation at large A  must be due to configurations further from those
of average cross section; in  particular the ones of relatively  small $\sigma$
(and therefore small size).
At the same time our numerical analysis shows that the series given by the
sum of cumulants of the cross
sections is badly convergent.
This is similar to the poor convergence of the standard Glauber series for
 $\sigma_{tot}(hA)$ which has an A-dependence of
$ A^{2/3}$ while the first term
$\sim A$.

In the limit that A becomes infinite, configurations of small size can
be expected to dominate since the nucleus acts as a black disk for  all
other configurations.
 Indeed Ref.~\cite{Bertsch81,ZKL81}
suggested that the effects of such small-sized
configurations would dominate the
pion-nucleus
 inelastic diffractive cross section.  This early result is inherent in our
eq. (2).
 The integration over small values of $\sigma$ in
eq. (2)  gives a result $\propto 1/T(B)$ for
 values of
 B within the nucleus.
 The integration over $d^2B$ leads to
 $\sigma_{diff}\propto R_A \propto A^{1/3}$.
Our realistic functions $P(\sigma)$ do not vanish at $\sigma=0$, so
we search for the dominance of small-sized configurations simply
by increasing A.
Numerical evaluations of eq.(2)  show that
the behavior is close to A$^{0.33}$ for fantastic  values of A
greater than about 10000.
An additional result
 indicating that small-sized configurations play a small role is
  obtained by simply cutting off the integrals over $\sigma$ at a maximum
value of
$\sigma_{max}= 5 mb$. This
 contribution varies from 2\% to 5\% as A increases from about 12 to 200.
This can be considered as an upper limit on the pQCD contribution suggested
in \cite{ZKL81,Bertsch81}.

The present result does not mean that small-sized configurations are
impossible to find. Reactions in which the pion diffractively
dissociates into two jets of high relative transverse momentum select those
configurations \cite{FMS93} and leads to an A$^2$ variation of the forward
diffractive
cross section. One could
 also look for the
 transition to this $A^2$ regime by considering
  production of
states where the value of
$M_{diff}$ is mainly determined
  by the transverse momenta of produced hadrons.

The same formalism and $P(\sigma)$ used to obtain
 $\sigma_{diff}(A)$ for pion and proton projectiles also
allow us to compute the A-dependence of the zero angle differential
cross section for coherent nuclear diffractive dissociation as well as
 the total
cross section.
The diffractive angular distribution
is related to the square of the scattering amplitude
${\cal M}
\sim \int d^2B
 e^{i\vec{q}_t\cdot \vec {B}}\langle h\mid F(\sigma,B)\mid X\rangle$.
Squaring $\cal M$ and summing over the diffractively
produced states X yields the angular distribution.
The result is
\begin{equation}
{d\sigma_{diff}\over dt}(0^\circ)
=\pi \int d\sigma P(\sigma) f^2(\sigma)-\pi\left[\int
 d\sigma P(\sigma) f(\sigma)\right]^2,
\end{equation}
where
$f(\sigma)\equiv \int B dB \left(1-e^{-{\sigma\over 2}T(B)}\right)$.
Numerical evaluation yields the result that
${d\sigma_{diff}\over dt}(0^\circ)$
varies approximately as
$A^{1.24}$ for the pion (see Fig 3) and A
for the proton. The total cross section    $\sigma_{tot}(A)$
is given  by the expression
\begin{equation}
\sigma_{tot}(A)= 2 \int d\sigma
P(\sigma)\int d^2B <h \mid F(\sigma,B) \mid h>
.\qquad
\end{equation}
Color fluctuations (also known in this case
 as inelastic shadowing \cite{Gri69})
have fairly small effects  on total cross sections. Numerical evaluation
 shows that the results of using the above equation are similar to
 those of
the more detailed calculations of Ref.~\cite{jm93}.

We also need to discuss the energy region for which our calculations are
valid. If  $E_h$ is much less than about 100 GeV, the cross
section $\sigma_{diff}(A)$ may increase significantly with energy due to the
 effects
of the nuclear longitudinal form factor $F_A$. Such effects are omitted here,
but are important in computing the inelastic shadowing correction to the total
nuclear cross section \cite{jm93} if
$E_h \le 100 GeV$.
In diffractive dissociation of a projectile h into a state of mass
M$_X$ the minimal longitudinal  momentum
$q_L$
transferred to the target is
$ (M_X^2-M_h^2)\over2 E_h$.
For small $q_L$ the nuclear  form factor can be described
using the  parameterization $F_A({\vec q}^{~2})=exp[-{\vec q}^{~2} R_A^2/3]$.
Thus if one uses $P(\sigma)$ for energies that vary as A$^{0.33}$
one can effectively use the same longitudinal form factor for different
nuclear targets.
This effect is a small correction if
$E_h$ is greater than about several hundred GeV.
But this is just the
energy range
where diffractive data and data on inelastic shadowing corrections
to
the total cross section of
hadron-deuteron scattering
are available.
A specific  energy dependence of $P(\sigma)$ expected at much higher energies
is discussed in ref. \cite{BBF92}.

Our results show that the effects of  color fluctuations
are important for coherent nuclear diffractive scattering.
 This has a broader significance because of its
relevance in heavy ion collisions.
In particular, the realistic $P(\sigma)$, used here to describe diffractive
processes, correspond to a significantly larger probability for
multiple scattering processes  to occur than the usual Glauber approximation.
Previous work \cite{HBB91}
 has shown that color fluctuations leads to significant
fluctuations of transverse energy in nucleus-nucleus collisions
in agreement with current data. Moreover the
work of Ref.~\cite {BF92} shows that
the probability for percolation phase transitions depends strongly on
the quantities $p_n(\sigma)
\equiv {\int_{\sigma}^\infty dy
y^n P(y)\over \langle \sigma^n \rangle}$.
These give the probability that n nucleon-nucleon inelastic collisions
occur with a cross section larger than $\sigma$. The differences between the
different versions of  $P(\sigma)$
are large and influence the predictions of whether or not a
percolation phase transition could occur in heavy ion collisions.
Measuring $\sigma_{diff}(A)$ for proton beams would strongly
constrain $P(\sigma)$.

Color fluctuations  have an important intrinsic interest through their
 close relation to QCD. New measurements of coherent nuclear diffraction
 could determine  finer details of $P(\sigma)$ and
 therefore
have a wide impact for studies of heavy ion collisions.
 We stress that new experiments
 are
 possible during the next fixed target run at FNAL  for a wide range of
 projectiles.

\begin {center} Acknowledgments
\end{center}

We thank V. Braun, T. Ferbel, B. Jennings, A. Mueller and D. Potter
for
useful discussions. This work was partially supported by the USDOE.

\newpage
\begin{center}
{\bf Figure Captions}
\end{center}

\noindent Figure 1a. $\sigma_{diff}(A)$, pion beam, for the different
$P(\sigma)$
of \cite{pion92}.
The data are from Ref. \cite{Ferbel1}.

\noindent Figure 1b. $\sigma_{diff}(A)$    for a proton beam.
The data are from Ref. \cite{Ferbel2}.

\noindent Figure 2.   $\sigma_{diff}(A)/\omega_\sigma$ (pion beam). Solid
curves from eq.(2).
Dashed curve- from eq.(4).

\noindent Figure 3. $\pi$ forward coherent nuclear diffraction
obtained using different $P(\sigma)$ of \cite{pion92}.
\end{document}